\documentstyle[12pt,epsfig]{article}
\begin{document}

\begin{center}
{\Large {\bf Quest for double beta decay of $^{160}$Gd and Ce isotopes}}
\end{center}

\vskip 0.7cm

\begin{center}
{\bf F.A.~Danevich, V.V.~Kobychev, O.A.~Ponkratenko,}

{\bf V.I.~Tretyak and Yu.G.~Zdesenko}\footnote{%
Corresponding author. Institute for Nuclear Research, Prospekt Nauki 47, MSP
03680 Kiev, Ukraine; fax: 380(44)265-4463; phone: 380(44)265-2210; e-mail:
zdesenko@kinr.kiev.ua}
\end{center}

\vskip 0.7cm

\begin{center}
{\it Institute for Nuclear Research, MSP 03680 Kiev, Ukraine}
\end{center}

\vskip 1.5cm

\begin{center}
{\bf Abstract}
\end{center}

\noindent
The $2\beta $ decay study of $^{160}$Gd has been performed in the Solotvina
Underground Laboratory with the help of Gd$_2$SiO$_5$:Ce crystal
scintillator (volume 95 cm$^3$). The background of the detector in the
vicinity of the $Q_{\beta \beta }$ energy of $^{160}$Gd was reduced to 1.0
cpd/keV$\cdot $kg. The new improved half-life limits have been established
for $0\nu 2\beta $ decay of $^{160}$Gd to the ground (0$^{+}$) and first
excited (2$^{+}$) levels of $^{160}$Dy: $T_{1/2}^{0\nu }\geq 2.3$(1.3)$%
\times 10^{21}$ yr at 68\%(90\%) C.L. The $T_{1/2}$ bounds have been also
set for $2\nu $, $0\nu \chi $ and $0\nu \chi \chi $ modes of $^{160}$Gd
decay, as well as for different $2\beta $ decay processes in $^{136}$Ce, $%
^{138}$Ce and $^{142}$Ce.

\vskip 0.5cm


\vskip 0.5cm


\section{Introduction}

An exceptional interest to the neutrinoless double beta ($0\nu 2\beta $)
decay is explained by the great potential of this process -- which violates
the lepton number conservation -- to search for the neutrino mass ($m_\nu $)
and its nature as a sign of a possible new physics beyond the standard model
(SM) \cite{Moe94,Tre95,Suh98,Fae98,Klap98,Vog00}. The absence of the $0\nu
2\beta $ decay, established at the required level of sensitivity, yields
strong restrictions on $m_\nu $, lepton violation constants and other
parameters of the manifold SM extensions, which allow to narrow a wide
choice of theoretical models and to touch the multi-TeV energy range
competitive to the accelerator experiments \cite{Suh98,Fae98,Klap98,Vog00}.

In fact, the most sensitive $0\nu 2\beta $ results were obtained by using
the so called ''active source'' technique, in which a detector (containing $%
2\beta $ candidate nuclei) serves as source and detector simultaneously \cite
{Moe94,Tre95}. As examples, we recall the impressive half-life limits $%
T_{1/2}^{0\nu }$ in the range of $(1-4)\times 10^{23}$ yr ($m_\nu \leq 1.3-4$
eV) established for $^{136}$Xe (high pressure Xe TPC) \cite{0nXe136}, $%
^{130} $Te (low temperature bolometers TeO$_2$) \cite{Te130}, $^{116}$Cd
(enriched $^{116}$CdWO$_4$ scintillators) \cite{Cd-00}, and the highest
limit reached for $^{76}$Ge (enriched HP $^{76}$Ge detectors): $%
T_{1/2}^{0\nu }\geq 1.8\times 10^{25}$ yr \cite{Ge76,IGEX}. Therefore, it is
apparent that application of the ''active source'' technique with a new
nucleus allows to extend the number of $2\beta $ candidates studied with a
high sensitivity\footnote{%
For example, the first interesting results were obtained recently for $%
2\beta $ decay processes in $^{40}$Ca and $^{46}$Ca with the help of newly
developed low radioactive CaF$_2$(Eu) crystal scintillators \cite{CaF2-99}.}.

During last years cerium-doped gadolinium silicate Gd$_2$SiO$_5$:Ce (GSO)
crystal scintillators have been developed \cite{GSO-90,Melch.-90}. These
scintillators are non-hygroscopic and have a large density (6.71 g/cm$^3$),
fast response (primary decay time about $30-60$ ns), quite high light output 
$(20\%$ of NaI(Tl), wavelength of emission maximum 440 nm). Moreover, it was
already demonstrated \cite{Gd-93,Kobay-95,Iwa97} that GSO crystals can be
applied for the $2\beta $ decay search of $^{160}$Gd, which is one of the
interesting candidate nucleus due to the following reasons. First, despite
rather low $2\beta $ decay energy release ($Q_{\beta \beta }=1729.7(13)$ keV 
\cite{Aud95}) its theoretical value of $T_{1/2}^{0\nu }\cdot \left\langle
m_\nu \right\rangle ^2=8.6\times 10^{23}$ yr is nearly three times lower
than that for $^{76}$Ge and $^{136}$Xe \cite{Staudt-90}, thus for the equal
measured $T_{1/2}^{0\nu }$ limits the experiment with $^{160}$Gd will yield
the more stringent restrictions on the neutrino mass and other parameters of
the theory. Secondly, recent calculation \cite{Castan-94} shows that
two-neutrino $2\beta $ decay of $^{160}$Gd is strongly forbidden due to
heavy deformation of this nucleus. Meanwhile, the suppression of the $0\nu
2\beta $ decay mode would be not so strong due to different sets of
intermediate states involved in both transitions. Therefore the energy
region of $0\nu 2\beta $ signal of $^{160}$Gd could be free of the
background from $2\nu 2\beta $ decays, which is very serious problem for $%
2\beta $ detectors with the poor energy resolution \cite{Tre95} . Thirdly,
the natural abundance of $^{160}$Gd is rather large (21.86\% \cite{Ros98})
allowing to build up the high sensitive apparatus with natural GSO crystals.

The present paper describes the new and further improved half-life limits on 
$2\beta $ decay of $^{160}$Gd obtained with the help of 95 cm$^3$ GSO
crystal scintillator and with about 3 times larger running time than in ref. 
\cite{Gd-96}, where preliminary results of this experiment have been already
published.

\section{Set up, background measurements and data analysis}

Cerium-doped gadolinium silicate crystal (5.4 cm long, 4.7 cm in diameter)
grown by Czochralski method was used in the measurements. The mass of the
crystal is 635 g, and the number of $^{160}$Gd nuclei is 3.951$\times
10^{23} $. The first 630 h of measurements had been carried out with the
mass of the crystal equal to 698 g, then its side surface had been ground on
1--1$.5$ mm.

The experiment was performed in the Solotvina Underground Laboratory (SUL)
of the INR in a salt mine 430 m underground ($\simeq $1000 m w. e., cosmic
muon flux $1.$7$\times $1$0^{-6}$ cm$^{-2}$ s$^{-1}$, neutron flux $\le $2.7$%
\times $1$0^{-6}$ cm$^{-2}$ s$^{-1}$, radon concentration in air $<$30 Bq m$%
^{-3}$) \cite{Zde88}. In the low background installation the GSO crystal is
viewed by the photomultiplier FEU-110 through a plastic light-guide 8.6 cm
in diameter and 18.2 cm long. The energy resolution of the detector was
measured in the energy region $60-2615$ keV by using $\gamma $ lines of $%
^{22}$Na, $^{137}$Cs, $^{207}$Bi, $^{226}$Ra, $^{232}$Th and $^{241}$Am
sources. As an example, the resolution equals $16.8\%$, $13.5\%$, $11.2\%$
and $10.7\%$ at the energy 662, 1064, 1770 and 2615 keV, respectively. In
course of measurement the energy calibration was carried out with $^{207}$Bi
source weekly. The passive shield made of high purity copper (5 cm
thickness), mercury (7 cm), and lead (15 cm) surrounds the GSO scintillator
to reduce the external background. Event-by-event data acquisition system
consists of the IBM PC compatible personal computer and CAMAC crate with
electronic units, which allow to record the amplitude (energy) and arrival
time of each event \cite{Dan95}.

Total statistics collected in the experiment is 13949 h ($1.015$ yr kg of
exposure). The measured background spectrum of the GSO crystal is depicted
in fig. 1, where the following peculiarities exist: the clear peak at the
energy 420 keV, the comparatively wide peak at the energy around 1050 keV
and two broad distributions dropped down at the energies 2.4 and 5.5 MeV.
Taking into account the relative light yield for $\alpha $ particles as
compared with that for electrons ($\alpha /\beta $ ratio) for the GSO
scintillator\footnote{%
The energy dependence of the $\alpha /\beta $ ratio was determined by using
the $\alpha $ peaks of $^{214}$Po, $^{215}$Po, $^{216}$Po and $^{220}$Rn
from the internal contamination of the crystal as following: $\alpha /\beta
=0.152+0.01765\cdot E_\alpha $, where $E_\alpha $ is in MeV. The mentioned
peaks were selected from the background with the help of the time-amplitude
analysis described below.}, the first peak is attributed to $\alpha $
particles of $^{152}$Gd ($T_{1/2}=1.0$8$\times $1$0^{14}$ yr; $E_\alpha
=2140 $ keV; abundance $\delta =0.20\%$) and $^{147}$Sm ($T_{1/2}=1.0$6$%
\times $1$0^{11}$ yr; $E_\alpha $ = $2233$ keV; $\delta =15\%$; samarium can
be present as impurity of the GSO crystal at the level of $\approx $8 ppm 
\cite{Gd-93}). The peak near 1050 keV as well as the broad distribution up
to the energy 2.4 MeV is mainly due to the radioactive contamination of the
crystal by the nuclides from the $^{232}$Th, $^{235}$U and $^{238}$U
families. The second distribution up to the energy 5.5 MeV is caused by
decays of $^{232}$Th daughter isotopes: a) $\beta $ decay of $^{208}$Tl ($%
Q_\beta =5.00$ MeV); b) $\beta $ decay of $^{212}$Bi ($Q_\beta =2.25$ MeV)
followed by $\alpha $ decay of its daughter $^{212}$Po ($T_{1/2}=0.3$ $\mu $%
s; $E_\alpha =8.78$ MeV or $\approx $2.7 MeV in $\beta $ scale).

\nopagebreak
\begin{figure}[ht]
\begin{center}
\mbox{\epsfig{figure=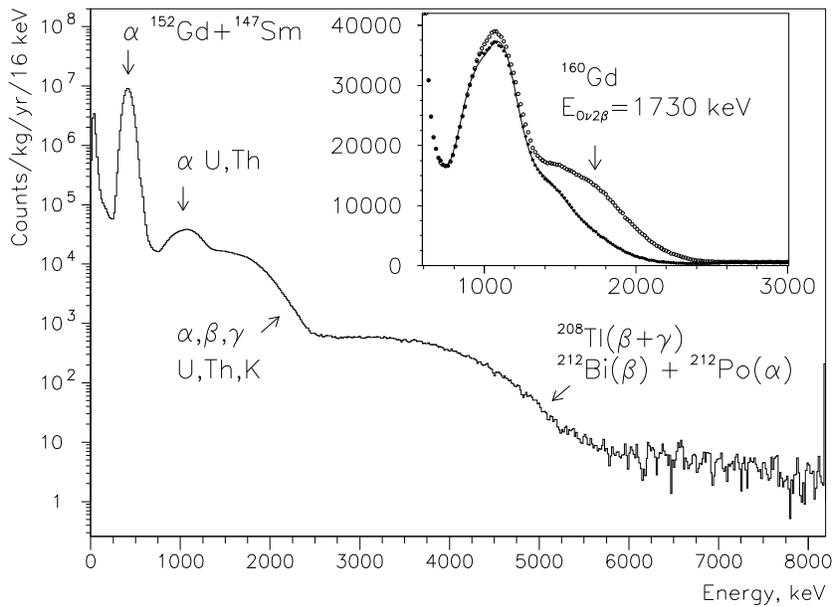,height=8.0cm}}
\caption {Background spectrum of the GSO crystal (95 cm$^3)$ collected
during 13949 h. In the insert: part of the measured distribution in the
energy interval 600 -- 3000 keV (open circles) together with the residual
after subtraction (see text) of the decays from the $^{226}$Ra, $^{227}$Ac
and $^{228}$Th intrinsic contamination of the crystal (filled circles) and
fitting curve (solid line).}
\end{center}
\end{figure}

To recognize and reduce the background from the intrinsic radioactive
impurities of the crystal, the off-line analysis of the arrival times of
measured events was fulfilled (as described in details elsewhere \cite
{Gd-96,Dan95}). Using this method fast sequences of decays belonging to the
natural radioactive chains were searched for, as for example the sequence of
two $\alpha $ decays from the $^{232}$Th family: $^{220}$Rn ($Q_\alpha =6.41$
MeV, ${\it T}_{1/2}=55.6$ s) $\rightarrow $ $^{216}$Po ($Q_\alpha =6.91$
MeV, $T_{1/2}=0.145$ s) $\rightarrow $ $^{212}$Pb. Because the energy of $%
^{220}$Rn $\alpha $ particles corresponds to 1.7 MeV in $\beta $ scale of
the GSO detector, the events within the energy region $1.2-2.2$ MeV were
used as triggers. Then all events following the triggers in the time
interval $10-1000$ ms (it contains the part $\eta _t$ = 0.945 of the total
number of the $^{216}$Po decays) were selected. As an example, the spectra
of the $^{220}$Rn and $^{216}$Po $\alpha $ decays obtained in this way -- as
well as the distribution of the time intervals between the first and second
events -- are presented in fig. 2. It is evident from this figure that
selected spectra and time distribution are in the excellent agreement with
those expected from $\alpha $ particles of $^{220}$Rn and $^{216}$Po. Taking
into account the efficiency of the time-amplitude analysis, the number of
accidental coincidences, and the interfered chain of $^{219}$Rn $\rightarrow
^{215}$Po (it is selected by the applied procedure with $2.04\%$
efficiency), the $^{228}$Th activity in the GSO crystal was determined as
2.287(13) mBq/kg. Then on the next step of analysis the fast couples found ($%
^{220}$Rn and $^{216}$Po) were used as triggers to search for preceding $%
\alpha $ decays of $^{224}$Ra ($Q_\alpha =5.79$ MeV, $T_{1/2}=3.66$ d). The
time window was set as $1-30$ s (it contains $\eta _t=0.30$ of $^{220}$Rn
decays). Resulting distribution which includes the accidental coincidences
(calculated ratio of the effect to the accidental background is equal to
0.825) is also in a good agreement with the expected $\alpha $ peak of $%
^{224}$Ra.

\nopagebreak
\begin{figure}[ht]
\begin{center}
\mbox{\epsfig{figure=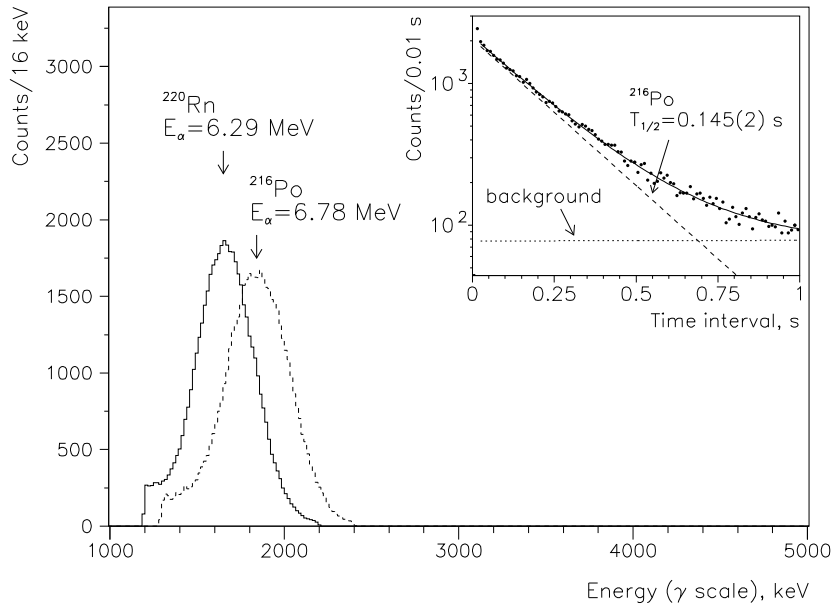,height=8.0cm}}
\caption {The energy spectra of the first and second $\alpha $
particles in the decay chain $^{220}$Rn $\rightarrow $ $^{216}$Po $%
\rightarrow $ $^{212}$Pb which were found with by means of the
time-amplitude analysis of the data recorded over 8609 h. In the insert: the
distribution of the time interval between the first and second events
together with its fit (solid line) by the sum of exponent (dashed line) with 
$T_{1/2}=0.145$ s (table value {\it T}$_{1/2}=0.145(2)$ s \cite{Fir96}) and
background (dotted line).}
\end{center}
\end{figure}

The same technique was applied to the sequences of $\alpha $ decays from the 
$^{235}$U family: $^{223}$Ra ($Q_\alpha =5.98$ MeV, $T_{1/2}=11.44$ d) $%
\rightarrow $ $^{219}$Rn ($Q_\alpha =6.95$ MeV, $T_{1/2}=3.96$ s) $%
\rightarrow $ $^{215}$Po ($Q_\alpha =7.53$ MeV, $T_{1/2}=1.78$ ms) $%
\rightarrow $ $^{211}$Pb. For the fast couple ($^{219}$Rn and $^{215}$Po),
events within $1.3-2.4$ MeV were used as triggers and the time interval of $%
0.5-10$ ms (containing $\eta _t=0.803$ of $^{215}$Po decays with energy
between 1.5 and 2.6 MeV) was chosen\footnote{%
The used procedure selects also pairs $^{220}$Rn $\rightarrow $ $^{216}$Po
and $^{214}$Bi $\rightarrow $ $^{214}$Po with efficiencies $4.64\%$ and $%
12.08\%$, respectively. These contributions are taken into account in the
calculation of activities.}. The obtained $\alpha $ peaks correspond to
activity of 0.948(9) mBq/kg for the $^{227}$Ac impurity in the crystal. Then
the procedure analogous to that described for the $^{224}$Ra was applied to
find the preceding $^{223}$Ra $\alpha $ decays.

For the analysis of the $^{226}$Ra chain ($^{238}$U family) the following
sequence of $\beta $ and $\alpha $ decays was used: $^{214}$Bi ($Q_\beta
=3.27$ MeV, $T_{1/2}=19.9$ m) $\rightarrow $ $^{214}$Po ($Q_\alpha =7.83$
MeV, $T_{1/2}=164.3$ $\mu $s) $\rightarrow $ $^{210}$Pb. For the first event
the lower energy threshold was set at 0.5 MeV, while for the second decay
the energy window $1.3-3.0$ MeV was chosen. Time interval of $2-500$ $\mu $s
($\eta _t=0.872$ of $^{214}$Po decays) was used. The selection efficiency is
also decreased a little by the energy threshold applied to the first event
(in the procedures described above the selection efficiency of the energy
windows $\eta _E$ is equal to $1$). By the Monte Carlo simulation the part $%
\eta _E$ of the $^{214}$Bi spectrum above 500 keV (as compared with the
total spectrum) has been determined as $\eta _E=0.794$. The obtained spectra
for $^{214}$Bi\footnote{%
Peak observed in the $\beta $ spectrum of $^{214}$Bi (see fig. 3) is the
part ($17.7\% $) of $^{219}$Rn $\alpha $ decays (from $^{235}$U family),
which corresponds to the chosen time interval $2 - 500$ $\mu $s. Due to
known activity of $^{227}$Ac the contribution of $^{219}$Rn $\rightarrow $ $%
^{215}$Po (as well as $0.24\% $ of the chain $^{220}$Rn $\rightarrow $ $%
^{216}$Po) was calculated accurately and subtracted from the activities of $%
^{214}$Bi and $^{214}$Po.} and $^{214}$Po are shown in fig. 3 and lead to
the $^{226}$Ra activity in the GSO crystal equal to 0.271(4) mBq/kg.

\nopagebreak
\begin{figure}[ht]
\begin{center}
\mbox{\epsfig{figure=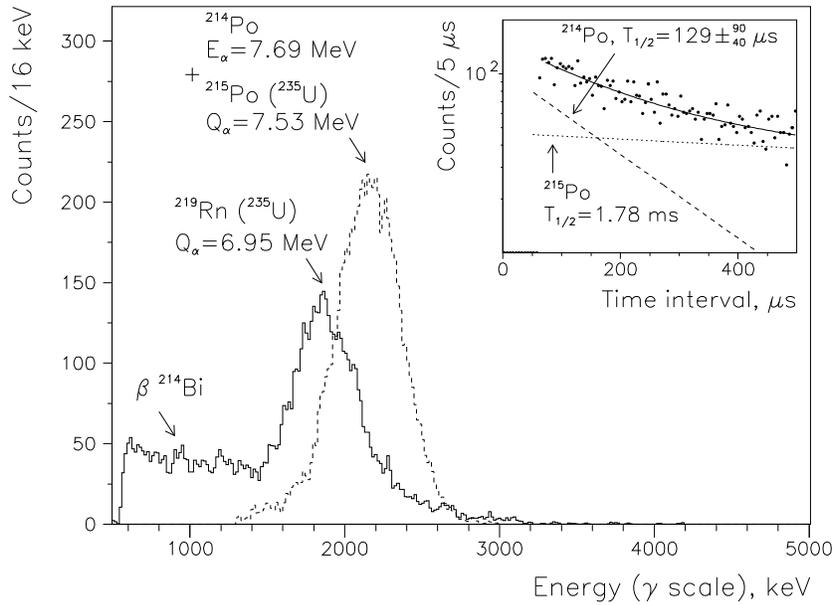,height=8.0cm}}
\caption {The energy spectra of the sequence of $\beta $ and $\alpha $
decays in the decay chain $^{214}$Bi $\rightarrow $ $^{214}$Po $\rightarrow $
$^{210}$Pb which were found by means of the time-amplitude analysis of 8609
h data. In the insert: the distribution of the time interval between the
first and second events together with its fit (solid line) by the sum of
exponent (dashed line) with $T_{1/2}=129$ $\mu $s (table value {\it T}$%
_{1/2}=164.3(20)$ $\mu $s \cite{Fir96}) and exponent with $T_{1/2}=1.78$ ms
related with the chain $^{219}$Rn $\rightarrow $ $^{215}$Po $\rightarrow $ $%
^{211}$Pb (dotted line).}
\end{center}
\end{figure}

Besides determination of the background components, the time-amplitude
analysis is used for reduction of the background. In this case, the
attention is paid, first, to selection of time-correlated decays with
minimal loss of the exposure, and, secondly, to prevention of the double
selection of the events. With these aims four ''removing'' procedures were
applied (the subscripts numerate events in the chain):

(i) $E_1\geq $ 500 keV; $\Delta t=2-500$ $\mu$s; $E_2=1000-3000$ keV.

This cut removes mainly decays of $^{214}$Bi and $^{214}$Po ($\eta _t=0.872$
, $\eta _{E_1}=0.794$, $\eta _{E_2}=1.00$, thus the total efficiency of
selection is $\eta =\eta _t\eta _{E_1}\eta _{E_2}=0.692$), while
efficiencies for the pairs of $^{219}$Rn -- $^{215}$Po and $^{220}$Rn --$%
^{216}$Po are rather low: $0.177$ and $0.024$, respectively.

(ii) $E_1$, $E_2=1200-3000$ keV; $\Delta t=500$ $\mu $s -- 1 s.

The corresponding efficiencies are: $\eta $($^{219}$Rn -- $^{215}$Po) $%
=0.823 $; $\eta $($^{220}$Rn -- $^{216}$Po) $=0.989$; and $\eta $($^{214}$Bi
-- $^{214}$Po) $=0.067$. It is also necessary to take into account the pairs
of $^{223}$Ra -- $^{219}$Rn ($\eta =0.161$) and $^{224}$Ra -- $^{220}$Rn ($%
\eta =0.0124$).

(iii) In the third cut a fast chain of decays ($^{220}$Rn -- $^{216}$Po) is
used as a trigger to select the previous event of $^{224}$Ra. The parameters
taken for $^{224}$Ra selection are the following: $E_1>$1000 keV; $\Delta
t_{12}=1-30$ s; $E_2=1200-2200$ keV; $\Delta t_{23}=$ $0.01-0.5$ s; $%
E_3=1300-2400$ keV. The final efficiency for selection of $^{224}$Ra is $%
\eta =0.258$.

(iv) In the fourth cut procedure a fast chain of decays ($^{219}$Rn -- $%
^{215}$Po) is used as a trigger to select the previous event of $^{223}$Ra.
The parameters are: $E_1>$1000 keV; $\Delta $$t_{12}=1-10$ s; $E_2=1300-2400$
keV; $\Delta t_{23}=500$ $\mu $s $-10$ ms; $E_3=1500-2600$ keV. It removes
53.4\% ($\eta =$ 0.$534)$ of $^{223}$Ra decays.

Since the time windows of the cuts are not overlapping, the probability that
one event could be selected more than once is rather low. Indeed, from the
total number of 345169 events being attributed to the time-correlated
background, only 318 (i. e. $0.092\%$) are double-selected.

The part of the measured distribution in the energy interval $600-3000$ keV
is depicted in the insert of fig. 1 (open circles) together with the
resulting spectrum after subtraction of the selected decays (filled
circles). It corresponds to $0.969$ yr$\cdot $kg of exposure ($\approx $
13400 h), or $95.7\%$ of the initial value that was decreased by the
''removing'' procedures. At the same time, due to these procedures the
background rate in the vicinity of the energy release of the $0\nu 2\beta $
decay of $^{160}$Gd ($1648-1856$ keV) has been reduced by 2.3 times up to
the value of 1.01(1) cpd/keV$\cdot $kg.

\section{Background simulation}

With the aim to evaluate the 2$\beta $ decay processes in $^{160}$Gd the
measured background spectrum (after subtraction of the time-correlated
decays) was simulated with the help of GEANT3.21 package \cite{GEA94}. The
event generator DECAY4 \cite{DECAY4} was used to describe initial kinematics
of decays (number and types of emitted particles, their energies, directions
of movement and times of emission). It takes into account decays to ground
state as well as to excited levels of daughter nuclei with the subsequent
complex de-excitation process \cite{Fir96}. The possibilities of emission of
conversion electrons and e$^{+}$e$^{-}$ pairs instead of $\gamma $ quanta in
nuclear transitions and the angular correlation between emitted particles
are also taken into consideration.

The background model was built up as a result of the procedure, in which the
experimental spectrum was fitted by the sum of simulated response functions.
The coefficients of the latest were determined on the basis of the following
data:

i) Activity values for $^{228}$Th, $^{227}$Ac and $^{226}$Ra (and their
short-lived daughters) that are present as intrinsic contamination in the
GSO crystal. These activities were determined firmly and accurately (with
uncertainty less than $1\%$) with the help of the time-amplitude analysis,
as described above. The part of decays of $^{224}$Ra, $^{220}$Rn, $^{216}$%
Po; $^{223}$Ra, $^{219}$Rn, $^{215}$Po; $^{214}$Bi and $^{214}$Po was
removed from the background spectrum, and the remaining part is calculated
with high precision for all these nuclides. Thus, one can describe the
spectrum of, for example, $^{228}$Th$+$daughters as a sum of the simulated
spectra with exactly known areas. If we suppose the secular equilibrium
within the natural radioactive chains for these contaminations, the
activities of the remaining long-lived members of the $^{232}$Th, $^{235}$U
and $^{238}$U families would be known too. However, it is known that
chemical procedures and crystal growth usually break the equilibrium in the
natural radioactive series. To account this possibility the activities of
the mentioned remaining members of the radioactive chains (namely, $^{232}$%
Th, $^{228}$Ra, $^{235}$U+$^{238}$U+$^{234}$U, $^{231}$Pa, $^{230}$Th, and $%
^{210}$Pb) were taken as free parameters for the fitting procedure\footnote{%
Three isotopes of uranium $^{235}$U, $^{238}$U and $^{234}$U are not
chemically separable, and their relative activities (0.046:1:1) do not
change, so we can take only one parameter for these nuclides.}.

ii) The radioactive impurities of the photomultiplier (PMT), which are the
main source of the external background. Their values for PMT FEU-110 were
measured previously \cite{Dan95} as 3.0(3) Bq ($^{40}$K), 0.8(2) Bq ($^{226}$%
Ra) and 0.17(7) Bq ($^{228}$Th). In the fit these activities were taken as
free parameters varied within their errors.

Besides, simulated spectra of $^{40}$K and $^{138}$La -- natural
radionuclides which could be present in the GSO crystal -- were included
into the fitting procedure too. The last component of the background model
is the exponential function (with two free parameters) which describes the
residual external background (multiple scattering of $\gamma $ quanta,
influence of weak neutron flux and so on). The exponential behavior of this
component was confirmed by the measurements with the high radiopurity CdWO$%
_4 $ crystal scintillator (454 g) performed in the same set up \cite{Cd-113}.

The fit of the experimental spectrum in the energy region $0.1-3.0$ MeV by
the sum of the described components gives the following activities of the
additional intrinsic contamination of the GSO crystal: $^{40}$K $\leq $ 14
mBq/kg; $^{138}$La $\leq $ 55 mBq/kg\footnote{%
This limit of $^{138}$La activity corresponds to the possible La impurity in
the GSO crystal at the level of 67 ppm, which does not contradict with the
results of the chemical analysis.}; $^{232}$Th $\leq $ 6.5 mBq/kg; $^{228}$
Ra $\leq $ 9 mBq/kg; $^{238}$U $\leq $ 2 mBq/kg; $^{231}$Pa $\leq $ 0.08
mBq/kg; $^{230}$Th $\leq $ 9 mBq/kg; $^{210}$Pb $\leq $ 0.8 mBq/kg. The
parameters of the exponent were also found in the fit procedure and its
contribution to the experimental distribution was revealed to be small ($%
\approx 2\%$ for the energy interval $1-2$ MeV). For illustration the
fitting curve (in the energy region $0.7-2.4$ MeV) is presented in the
insert of fig. 1 together with the experimental data.

\section{Half-life limits on the 2$\beta $ decay of $^{160}$Gd}

Since in the measured spectrum the $0\nu 2\beta $ decay peak of $^{160}$Gd
is evidently absent, only the limit for the probability of this process can
be set on the base of the experimental data. To estimate the half-life limit 
$T_{1/2}$, we use the formula lim\quad $T_{1/2}=\ln 2\cdot \eta \cdot N\cdot
t/\lim S$,\quad where $\eta $ is the detection efficiency, $N$ is the number
of $^{160}$Gd nuclei, $t$ is the measuring time and $\lim S$ is the number
of effect's events which can be excluded with a given confidence level. To
calculate the values of $\eta $ and $\lim S$, the response function of the
GSO detector for the effect being sought has been simulated with the help of
GEANT3.21 and DECAY4 programs. It was found that for the $0\nu 2\beta $
decay the response function is the Gaussian centered at 1730 keV and its
width (FWHM) equals 176 keV. The edge effects (escape of one or both
electrons and bremsstrahlung quanta from the crystal) remove from the peak $%
\approx 5\%$ of events, thus $\eta =0.95$. The $\lim S$ values were
determined in two ways. First, by using the so called ''one $\sigma $
approach'', in which the excluded number of effect's events is estimated
simply as square root of the number of background counts in a suitably
chosen energy window $\Delta E$. Notwithstanding its simplicity this method
gives the right scale of the sensitivity of the experiment. For instance, in
the measured spectrum within the energy interval $1648-1856$ keV (it
contains $82\%$ of the expected peak area) there are 74500 counts; thus, the
square root estimate gives $\lim S=273$ events. Using this value of $\lim S$
, the total exposure related to $^{160}$Gd nuclei ($N\cdot t=6.04\times
10^{23}$ nuclei yr), and the calculated efficiency ($\eta =0.78$), we obtain
the half-life limit: {\it T}$_{1/2}\geq 1.2\times 10^{21}$ yr $($68\% C.L.$)$%
. Further, the $\lim S$ value was determined by using the standard least
squares procedure, where the experimental energy distribution in the
vicinity of the peak searched for was fitted by the sum of the background
model (as described above) and effect's peak being sought. It should be
stressed that for the energy interval of interest the main important
contributions ($\approx $73$\%$ of the experimental spectrum within 1600 --
1900 keV) are the activities of $^{226}$Ra, $^{227}$Ac and $^{228}$Th (and
their short-lived daughters) from the intrinsic contamination of the GSO
crystal, whose values were determined accurately (better than $1\%$
uncertainty).

As a result of the fit procedure in the energy region 1.3--2.1 MeV, the
obtained area for the $0\nu 2\beta $ decay peak is $-160\pm 233$ counts ($%
\chi ^2$ value equals to 0.85), thus giving no evidence for the effect. The
difference between measured and simulated spectra in the energy region of
the hypothetical $0\nu 2\beta $ decay peak of $^{160}$Gd is shown in fig. 4,
where the excluded effect is also depicted (solid line). The number of
effect's events, which can be excluded with 90\%(68\%) C.L. was calculated 
\cite{PDG96} as 298(169). It gives the half-life limit:

\nopagebreak
\begin{figure}[ht]
\begin{center}
\mbox{\epsfig{figure=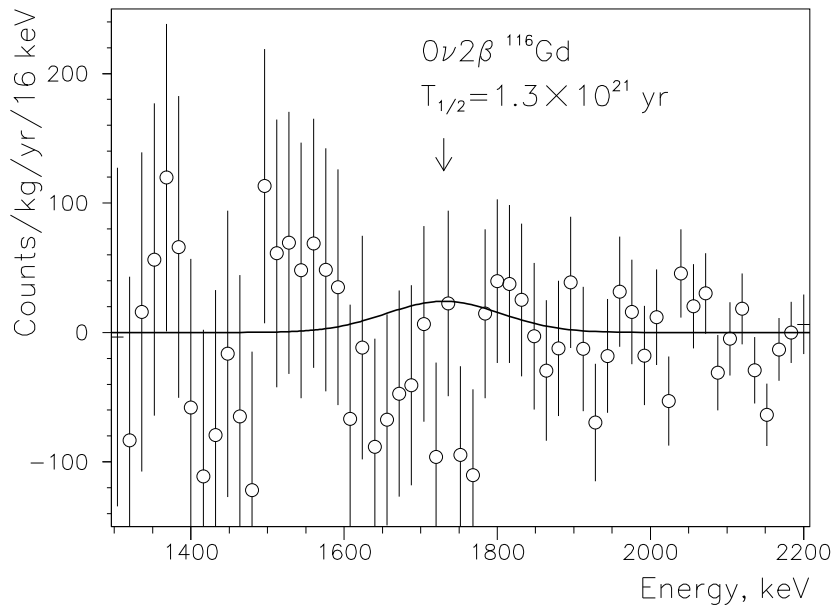,height=8.0cm}}
\caption {The residual between the experimental and simulated spectra
in the energy region of the $0\nu 2\beta $ decay of $^{160}$Gd (open circles
with error bars). The solid line represents the excluded $0\nu 2\beta $ peak
with $T_{1/2}^{0\nu }=1.3\times 10^{21}$ yr (90\% C.L.).}
\end{center}
\end{figure}

\begin{center}
$T_{1/2}$(0$\nu $2$\beta )\geq 1.3(2.3$)$\times $1$0^{21}$ yr \qquad at
90\%(68\%) C.L.
\end{center}

Comparing our limit with the theoretical calculations \cite{Staudt-90}, one
can compute the restriction on the neutrino mass $\left\langle m_\nu
\right\rangle \leq $ 26(19) eV at 90\%(68\%) C.L.

The same $T_{1/2}$ limit has been set for the $0\nu 2\beta $ transition to
the first excited level of $^{160}$Dy (2$^{+}$, 87 keV) because 87 keV $%
\gamma $ quanta following this process will be almost fully absorbed inside
the scintillator. The obtained limits are several times larger than those
already published \cite{Gd-93,Kobay-95,Iwa97,Gd-96}.

The two neutrino $2\beta $ decay rate was evaluated in two ways. For the
very conservative estimate the model spectra of the exactly measured
contamination in the crystal and PMT were subtracted from the experimental
data. The residual in the chosen energy region was equated to the simulated $%
2\nu 2\beta $ decay distribution and the latest was taken as an excluded
effect with $\lim S=$ 7.6$\times 10^5$ events corresponding to the $%
T_{1/2}^{2\nu }$(g.s.) $\geq 5.$5$\times $1$0^{17}$ yr ($99\%$ C.L.) for $%
2\nu 2\beta $ decay of $^{160}$Gd. However, within this simple approach it
was impossible to reproduce adequately the measured spectrum, thus the
background model and the fitting procedure described above were applied for
the $2\nu $ decay mode too. The set of fits were performed by changing the
fitting energy region from (100 $-$ 760) keV to (2400 -- 3000) keV. The
maximum value of an excluded effect ($\lim S=$ 2.2$\times $1$0^4$ events at $%
90\%$ C.L.) was found for the energy interval 760 -- 2600 keV. The
corresponding Monte Carlo simulated spectrum of $2\nu 2\beta $ decay of $%
^{160}$Gd (g.s.$\rightarrow $g.s.) is shown in fig. 5 together with the
fitting curve and the most important background components. It is visible
from this figure that our background model reproduces the experimental data
quite well even outside the region of the fit. The final lower limit for the
process searched for is equal to:

\nopagebreak
\begin{figure}[ht]
\begin{center}
\mbox{\epsfig{figure=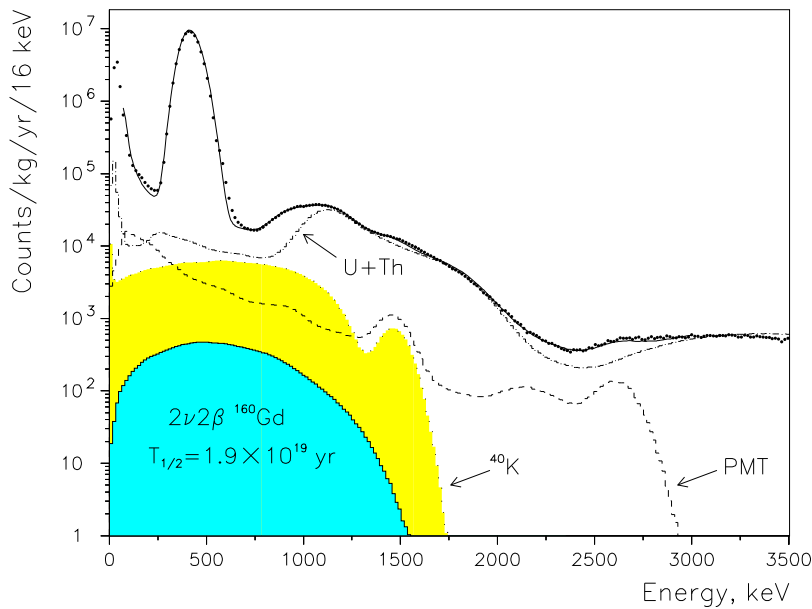,height=8.0cm}}
\caption {The background spectrum of the GSO detector for $0.969$ yr$%
\cdot $kg of exposure (points) and the model of background (solid line)
obtained by the fitting procedure in $760-2600$ keV energy interval (see
text). The most important internal ($^{40}$K and sum of $^{238}$U, $^{235}$%
U, $^{232}$Th) and external ($\gamma $ radiation from PMT) components of
background are shown. The excluded with $90\%$ C.L. distribution of $2\nu
2\beta $ decay of $^{160}$Gd to the ground level of $^{160}$Dy corresponds
to the half-life limit $T_{1/2}^{2\nu }$(g.s.) $=1.9\times 10^{19}$ yr.}
\end{center}
\end{figure}

\begin{center}
$T_{1/2}^{2\nu }$(g.s.) $\geq 1.9(3.1)\times 10^{19}$ yr \qquad at $90\%$($%
68 \%$) C.L.
\end{center}

The same method gives the limit for $2\nu 2\beta $ transition to the first (2%
$^{+}$) excited level of $^{160}$Dy\footnote{%
It should be noted that shape of distribution for $2\nu 2\beta $ decay to 2$%
^{+}$ level differs from $2\nu 2\beta $ transition to ground (0$^{+}$) state 
\cite{Tre95}.}:

\begin{center}
$T_{1/2}^{2\nu }$(2$^{+}$) $\geq 2.1(3.4)\times 10^{19}$ yr \qquad at $90\%$(%
$68 \%$) C.L.
\end{center}

The similar fitting procedure was used to set limits on double beta decays
with one or two Majoron emission. The estimated restrictions on half-life
are:

\begin{center}
$T_{1/2}^{0\nu \chi }\geq 3.5(5.3)\times 10^{18}$ yr\qquad at $90\%$($68 \%$%
) C.L.

$T_{1/2}^{0\nu \chi \chi }\geq 1.3(2.0)\times 10^{19}$ yr\qquad at $90\%$($%
68 \%$) C.L.
\end{center}

\section{Limits on 2$\beta $ decay processes of Ce isotopes}

The concentration of cerium in the GSO(Ce) crystal ($0.8\%$) is known from
the crystal growth conditions and from the results of the chemical analysis.
It allows to search for the $2\beta $ processes in three cerium isotopes:
double positron decay (2$\beta ^{+}$), or electron capture and positron
decay ($\epsilon \beta ^{+}$), or double electron capture (2$\epsilon $) in $%
^{136}$Ce (mass difference between parent and daughter atoms $\Delta M_A$ =
2397(48) keV; abundance of parent nuclide $\delta =$ $0.185\%$); double
electron capture in $^{138}$Ce ($\Delta M_A=693(11)$ keV; $\delta =0.251\%$
); and 2$\beta ^{-}$ decay in $^{142}$Ce ($\Delta M_A=1417(2)$ keV; $\delta
=11.114\%$). The total numbers of $^{136}$Ce, $^{138}$Ce and $^{142}$Ce
nuclei in the crystal are 4.1$\times 10^{19}$, 5.4$\times 10^{19}$ and 2.4$%
\times 10^{21}$, respectively. The response functions of the detector for
the different possible $2\beta $ processes in these cerium isotopes were
simulated with the help of GEANT3.21 code and event generator DECAY4. As an
example, the simulated energy distributions for neutrinoless and
two-neutrino 2$\beta ^{+}$ (K$\beta ^{+}$and 2K) decays of $^{136}$Ce are
presented in fig. 6. Limits on half-lives with respect to different modes of
2$\beta $ processes of cerium isotopes were calculated in the way analogous
to that for $^{160}$Gd -- they are summarized in the Table. Except two
limits for 0$\nu $2$\beta ^{+}$ decay of $^{136}$Ce and 0$\nu $2$\beta ^{-}$
decay of $^{142}$Ce (already obtained by using CeF$_3$ scintillators \cite
{Ber97}), all presented results for cerium isotopes are set for the first
time.

\nopagebreak
\begin{figure}[ht]
\begin{center}
\mbox{\epsfig{figure=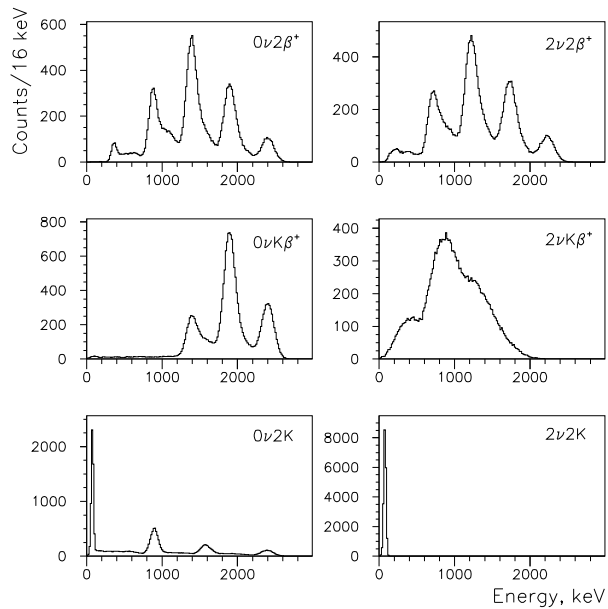,height=8.0cm}}
\caption {The response functions of the GSO(Ce) detector for
neutrinoless and two-neutrino 2$\beta ^{+}$ (K$\beta ^{+}$ and 2K) decays of 
$^{136}$Ce simulated with the help of GEANT3.21 code and event generator
DECAY4.}
\end{center}
\end{figure}

\section{Conclusions}

The present experiment performed in the Solotvina Underground Laboratory
with the help of GSO scintillator has shown ''pro et contra'' of these
crystals for the advanced double $\beta $ decay study of $^{160}$Gd.

The current level of sensitivity is limited mainly by the intrinsic
contamination of the GSO crystal ($^{232}$Th, $^{235}$U and $^{238}$U
families), therefore further development (which is in progress now) to
remove these impurities from the crystals is at most importance. Careful
purification of raw materials from actinides and their daughters
(technically available now) could decrease radioactive contaminations of the
GSO crystals by two-three orders of magnitude down to the level of several $%
\mu $Bq/kg\footnote{%
Such a radiopurity has been already reached in the CdWO$_4$ crystal
scintillators used for the $2\beta $ decay study of $^{116}$Cd \cite
{Dan95,PTE,Bur96}.}. Another possibility to reduce background is the
pulse-shape discrimination. Our preliminary measurement shows the difference
of decay times for pulses induced in GSO by $\alpha $ particles and $\gamma $
quanta: 50.8(4) and 48.3(4) ns, respectively\footnote{%
These results are in a good agreement with ref. \cite{Ish89}.}. Although
this difference is not so large, the proper pulse shape discrimination
technique, which is under development now, would allow to eliminate $\alpha $
background from decays of $^{147}$Sm, $^{152}$Gd and members of $^{232}$Th, $%
^{235}$U and $^{238}$U families.

On the other hand, due to the high abundance of $^{160}$Gd ($\approx 22\%$)
the GSO crystals could be grown up from natural Gd, thus such detectors
should be well less expensive than those made of enriched $2\beta $ decay
candidate isotopes. Therefore, the enforcement of the large scale and high
sensitivity experiment with $^{160}$Gd would be really possible by using the
GSO multi-crystal array with total mass of about 2000 kg ($\approx $400 kg
of $^{160}$Gd).

The background of such a detector could be reduced further by placing these
crystals into a high purity liquid (water or scintillator)\footnote{%
The existing and future large underground neutrino detectors (SNO \cite{SNO}
, BOREXINO \cite{BOREXINO}, KamLand \cite{KamLand}) could be appropriate for
our proposal with one-two tons of the GSO crystals. The latest located in
the water or liquid scintillator would be homogeneously spread out on a
sphere with diameter 3--4 m and viewed by the distant PMTs.} serving as
shield and light guide simultaneously. The similar idea has been recently
proposed in CAMEO project \cite{CAMEO} with aim to study 2$\beta $ decay of $%
^{116}$Cd by using $^{116}$CdWO$_4$ crystals placed in the liquid
scintillator of the BOREXINO Counting Test Facility (CTF). It was evidently
demonstrated by pilot measurements with $^{116}$Cd \cite{Cd-00} and by Monte
Carlo simulation that sensitivity of the CAMEO experiment (in term of the $%
T_{1/2}$ limit for $0\nu 2\beta $ decay) with $\approx $ 100 kg of $^{116}$%
CdWO$_4$ crystals is $\approx $10$^{26}$ yr which translates to the
constraint on the neutrino mass $m_\nu \leq 0.06$ eV \cite{CAMEO}. Moreover,
the strong dependence of the light collected by each PMT versus coordinate
of the emitting source in the crystal has been found. Such a dependence is
explained by difference of the refraction indexes of crystal ($n$ = $2.3$
for CdWO$_4$) and liquid scintillator ($n$ = $1.58$), which leads to the
redistribution between reflected and refracted light due to change of the
source position. By means of the GEANT Monte Carlo simulation it was shown
that spatial resolution of 1-- 5 mm (depending on the event location and the
energy deposit) can be reached with CdWO$_4$ crystals ($\oslash 7\times 9$
cm) placed in the liquid scintillator of the CTF and viewed by 200 distant
PMTs \cite{CAMEO}. With the GSO detectors (refractive index $n$ = $1.85$)
placed in a liquid ($n$ $\approx $ $1.5$), the simulation of light
propagation gives the spatial resolution in the range of 4 -- 10 mm. Anyhow,
it would certainly allow to reduce background in the energy region of
interest additionally (roughly by factor of 10 -- 50).

We estimate that sensitivity of the experiment with about two tons of the
GSO crystals (placed in the SNO or BOREXINO set ups) and for 5 -- 10 years
of exposition would be of the order of $T_{1/2}^{0\nu }\approx $2$\times $10$%
^{26}$ yr, hence the restriction on the Majorana neutrino mass can be
reduced down to $m_\nu \leq 0.07$ eV. It is comparable with the
sensitivities of the recently proposed large scale projects for $2\beta $
decay study, like MOON \cite{MOON}, EXO \cite{EXO}, CAMEO \cite{CAMEO},
CUORE \cite{CUORE}, MAJORANA \cite{MAYOR}, GENIUS \cite{GENIUS}, whose
results could provide crucial tests of the certain key problems and
theoretical models of the modern astroparticle physics. Note, however, that
cost of the GSO experiment would be well lower than those of mentioned
projects.

The authors would like to thank A.Sh.~Georgadze, B.N.~Kropivyansky,
A.S.~Nikolaiko, and S.Yu.~Zdesenko for their participation in the
measurements.

\newpage

\begin{table}[tbp]
\caption{Half-life limits on the $2\beta $ decay processes in $^{160}$Gd, $%
^{136}$Ce, $^{138}$Ce and $^{142}$Ce}
\begin{center}
\begin{tabular}{|l|lll|ll|}
\hline
Nuclide & \multicolumn{3}{c|}{Decay mode} & \multicolumn{2}{c|}{Limit on T$%
_{1/2}$, yr} \\ 
~ & ~ & ~ & ~ & present work & other works, (C.L.) \\ 
~ & ~ & ~ & ~ & 90(68)\% C.L. & ~ \\ \hline
~ &  &  &  &  &  \\ 
$^{160}$Gd & 2$\beta^-$ & $0\nu$ & g.s.--g.s., $2^+$ & $1.3(2.3)\times
10^{21}$ & $1.4\times 10^{19}$ (90\%) \cite{Gd-93} \\ 
~ & ~ & ~ & ~ & ~ & $3.0\times 10^{20}$ (68\%) \cite{Kobay-95} \\ 
~ & ~ & ~ & ~ & ~ & $8.2\times 10^{20}$ (90\%) \cite{Gd-96} \\ 
~ & ~ & $2\nu$ & g.s.--g.s. & $1.9(3.1)\times 10^{19}$ & $1.3\times 10^{17}$
(99\%) \cite{Gd-93} \\ 
~ & ~ & $2\nu$ & g.s.$-2^+$ & $2.1(3.4)\times 10^{19}$ & ~ \\ 
~ & ~ & $0\nu \chi$ & g.s.--g.s. & $3.5(5.3)\times 10^{18}$ & $2.7\times
10^{17}$ (99\%) \cite{Gd-93} \\ 
~ & ~ & $0\nu \chi \chi $ & g.s.--g.s. & $1.3(2.0)\times 10^{19}$ & ~ \\ 
~ &  &  &  &  &  \\ 
$^{136}$Ce & 2$\beta^+$ & $0\nu$ & g.s.--g.s. & $1.9(3.2)\times 10^{16}$ & $%
6.9\times 10^{17}$ (68\%) \cite{Ber97} \\ 
~ & ~ & $2\nu$ & g.s.--g.s. & $1.8(3.8)\times 10^{16}$ & ~ \\ 
~ & K$\beta^+$ & $0\nu$ & g.s.--g.s. & $3.8(6.0)\times 10^{16}$ & ~ \\ 
~ & ~ & $2\nu$ & g.s.--g.s. & $1.8(3.0)\times 10^{15}$ & ~ \\ 
~ & 2K & $0\nu$ & g.s.--g.s. & $6.0(8.0)\times 10^{15}$ & ~ \\ 
~ & ~ & $2\nu$ & g.s.--g.s. & $0.7(1.1)\times 10^{14}$ & ~ \\ 
~ &  &  &  &  &  \\ 
$^{138}$Ce & 2K & $0\nu$ & g.s.--g.s. & $1.8(1.9)\times 10^{15}$ & ~ \\ 
~ & ~ & $2\nu$ & g.s.--g.s. & $0.9(1.5)\times 10^{14}$ & ~ \\ 
~ &  &  &  &  &  \\ 
$^{142}$Ce & 2$\beta^-$ & $0\nu$ & g.s.--g.s. & $2.0(3.3)\times 10^{18}$ & $%
1.5\times 10^{19}$ (68\%) \cite{Ber97} \\ 
~ & ~ & $2\nu$ & g.s.--g.s. & $1.6(2.6)\times 10^{17}$ & ~ \\ 
~ &  &  &  &  &  \\ \hline
\end{tabular}
\end{center}
\end{table}

\end{document}